\documentclass[useAMS,usenatbib]{mn2e}

\usepackage{caption}
\usepackage{graphicx}
\usepackage{textcomp}
\usepackage{latexsym}
\usepackage{varioref}
\usepackage{xspace}
\usepackage{makeidx}
\usepackage{verbatim}
\usepackage{tabularx}
\usepackage{epstopdf}
\usepackage{amsmath}
\usepackage{array}
\usepackage{rotating}
\usepackage{float}

\def\Kepler{\textit{Kepler}}

\title[Fourier-dependent time lags in CVs]
{Discovery of Fourier-dependent time lags in cataclysmic variables}
\author[S. Scaringi \textit{et al.}]
{S. Scaringi$^{1,2}$\thanks{E-mail: simone.scaringi@ster.kuleuven.be}, E. K\"{o}rding$^{2}$, P.J. Groot$^{2}$, P. Uttley$^{3}$, T.Marsh$^{4}$, \newauthor C. Knigge$^{5}$, T. Maccarone$^{5,6}$, V.S. Dhillon$^{7}$\\ 
$^{1}$Instituut voor Sterrenkunde, K.U. Leuven, Celestijnenlaan 200D, B-3001 Leuven, Belgium \\
$^{2}$Department of Astrophysics/IMAPP, Radboud University Nijmegen, P.O. Box 9010, 6500 GL Nijmegen, The Netherlands \\ 
$^{3}$Astronomical Institute ``Anton Pannekoek'', University of Amsterdam, Science Park904, 1098 XH, Amsterdam, The Netherlands \\ 
$^{4}$Department of Physics, Astronomy and Astrophysics group, University of Warwick, CV4 7AL Coventry, UK  \\ 
$^{5}$School of Physics and Astronomy, University of Southampton, Highfield, Southampton, SO17 1BJ, UK \\ 
$^{6}$Astronomy \& Astrophysics group, Texas Tech University, Lubbock, TX 79409-1051, USA  \\ 
$^{7}$Department of Physics and Astronomy, University of Sheffield, Sheffield, S3 7RH, UK \\ 
}
\begin{document} 

\date{}

\pagerange{\pageref{firstpage}--\pageref{lastpage}} \pubyear{2013}

\maketitle

\label{firstpage}

\begin{abstract}
We report the first study of Fourier-frequency-dependent coherence and phase/time lags at optical wavelengths of cataclysmic variables (MV Lyr and LU Cam) displaying typical flickering variability in white light. Observations were performed on the William Herschel Telescope using ULTRACAM. Lightcurves for both systems have been obtained with the SDSS filters $u'$, $g'$ and $r'$ simultaneously with cadences between $\approx0.5-2$ seconds, and allow us to probe temporal frequencies between $\approx10^{-3}$ Hz and $\approx1$ Hz. We find high levels of coherence between the $u'$, $g'$ and $r'$ lightcurves up to at least $\approx10^{-2}$ Hz. Furthermore we detect red/negative lags where the redder bands lag the bluer ones at the lowest observed frequencies. For MV Lyr time lags up to $\approx3$ seconds are observed, whilst LU Cam displays larger time lags of $\approx10$ seconds. Mechanisms which seek to explain red/negative lags observed in X-ray binaries and Active Galactic Nuclei involve reflection of photons generated close to the compact object onto the surface layers of the accretion disk, where the lag delay is simply the light-travel time from the emitting source to the reflecting accretion disk area. Although this could be a viable explanation for the lags observed in MV Lyr, the lags observed in LU Cam are too large to be explained by reflection from the disk and/or the donor star. We suggest reprocessing on the thermal timescale of boundary layer photons onto the accretion disk as a possible mechanism to explain the lags observed in accreting white dwarfs, or reverse (inside-out) shocks within the disk travelling through cooler disk regions as they move outwards.
\end{abstract}


\section{Introduction}

Cataclysmic variables (CVs) are close interacting binary systems where a late-type star transfers material to a white dwarf (WD) companion via Roche lobe overflow. With a system orbital period ranging from hours to minutes, the transferred material from the secondary star forms an accretion disk surrounding the WD. As angular momentum is transported outwards in the disk, material will approach the inner-most regions close to the WD in the absence of strong magnetic fields, and eventually accrete onto the compact object. X-ray binaries (XRBs) are also compact interacting binaries which are similar to CVs in many ways, but where the accreting compact object is either a black hole (BH) or a neutron star (NS). Both CVs and XRBs, as well as Active Galactic Nuclei (AGN; accreting extragalactic supermassive black holes) have been shown to display aperiodic variability on a broad range of timescales, with often very strong fractional rms amplitude. XRBs have shown variability ranging from milliseconds to hours, whilst for CVs this ranges from seconds to days. This difference can be mainly attributed to the fact that the innermost edges of the accretion disks in CVs sit at a few thousand gravitational radii ($r_g$), whilst for XRBs it can reach down to just a few $r_g$. The fact that material can get deeper within the gravitational potential of XRBs, as compared to CVs, also explains why they are more luminous and emit predominantly in X-rays, compared to CVs, which emit predominantly at optical/UV wavelengths.

Aperiodic broad-band variability (also referred to as flickering) has been extensively studied in X-rays for XRBs over several decades in temporal frequency (see for example \citealt{terrell72,vanderklis_aper,belloni1,belloni2,homan}). As CVs emit mostly at optical/UV wavelengths, timing studies of these objects had to rely on optical observing campaigns from Earth, which are inevitably hindered by large interruptions, poor cadence, and in many cases poor signal-to-noise ratios. Furthermore, the key timescales to probe in CVs are much longer than in XRBs, requiring long, uninterrupted observations. Recently CV timing studies have been facilitated thanks to the advent of the \Kepler\ satellite (\citealt{jenkins,gilliland}), which is able to provide long, uninterrupted and high precision lightcurves in the optical light from space. Thanks to these capabilities it is now possible to probe over four orders of magnitude in temporal frequency in CVs. One important discovery in this respect is that the flickering properties of CVs in the optical are very similar, at least phenomenologically, to those observed in X-rays for XRBs (\citealt{scaringi_rms,scaringi_qpo,warner03,mauche}). More specifically, the discovery of linear rms-flux relations in the CV MV Lyr (as well as in XRBs and AGN,\citealt{uttley1,uttley2}) seems to rule out simple additive processes as the source of flicker noise (e.g. superposition of many independent ``shots''), and instead strongly favoured multiplicative processes (e.g. mass-transfer variations travelling from the outer to inner disc for the latter) as the source of variability.

Additional similarities between the broad-band timing properties of XRBs and AGN are observed in X-rays when analysing simultaneous lightcurves obtained in two energy bands. Frequency-dependent phase/time lags are detected for both XRBs and AGN over a wide range of frequencies (\citealt{fabian09,nowak,vaughan,demarco,uttley11,cassatella1,cassatella2}), where X-ray hard/blue photons lag X-ray soft/red ones at low frequencies (by $\approx$ hours for AGN and by $\approx$ seconds for XRBs), and where the opposite is observed at higher frequencies (by $<1$ hour for AGN and by $<1$ second for XRBs). Additionally, in some XRBs, the X-ray phase lags are observed to change sign as a function of intensity,hardness and/or frequencies (\citealt{reig}). The reason for these Fourier-dependent phase/time lags is still debated, but viable scenarios exist to explain the observed phenomena. For example, the fluctuating accretion disk model (\citealt{lyub,arevalo}) assumes fluctuations in the mass-transfer rate within the disk to cause the observed variability over a wide range of frequencies, with long timescale variability being produced further out in the accretion disk as compared to the fast timescale variability produced further in. As the long timescale variations propagate inwards, they couple to the fast timescale variations, which also explains the observed rms-flux relations (\citealt{uttley1,scaringi_rms,heil}). A consequence of this model is that we should observe blue/hard photons lagging redder/softer ones at the longest observable frequencies as a consequence of fluctuations propagating inwards and thus passing through different (and hotter) emitting regions in the disk. On the other hand, the soft/negative lag (where soft photons lag hard ones) should be observed at the highest frequencies, and are explained by reprocessing of X-ray photons produced close to the compact object (by the Comptonised component observed as a power law in hard X-ray spectra) onto the accretion disk, either as a thermal blackbody (e.g. X-ray heating from the disk) or, in the case of AGN, as an additional soft photoionised reflection component. In this case the size of the negative lags would provide an indication of the size of the reprocessing region in the disk: hard/blue photons are seen first as they are reprocessed closer in the accretion disk than the soft/red ones, and the time lag would simply be the light travel time from the central compact object to the reprocessing region plus a small reprocessing time.

A comprehensive analysis of coherence, phase and time lags at optical wavelengths has never been performed, and because of this no such study has ever been attempted for CVs\footnote{Although instruments such as ULTRACAM allow for fast, multi-colour and simultaneous photometry, and has been operational for several years.}. In this paper we present the first analysis with data obtained on the 4.2 meter William Herschel Telescope, using ULTRACAM, to study Fourier-dependent coherence, phase and time lags over three orders of magnitude in temporal frequency for two CVs, LU Cam and MV Lyr. This analysis provides useful insight into the broad-band variability behaviour of CVs, and reveals further similarities to the behaviour observed in X-rays for XRBs and AGN. 

MV Lyr is classified as being a VY Scl novalike system, spending most of its time in a high state (V$\approx$ 12-13), but occasionally (every few years) undergoing short-duration (weeks to months) drops in brightness (V$\approx$16-18, \citealt{hoard}). The reason for these sudden drops in luminosity is not ccd ..
rm Lag	lear, but suggestions involving star spots on the secondary inhibiting mass transfer have been proposed (\citealt{LP}). An orbital period of 3.19 hours has been determined for the system, as well as a low inclination of $i\approx 11^o - 13^o$ (\citealt{SPT95}). \cite{scaringi_rms,scaringi_qpo} have already studied the single-band variability behaviour using \Kepler\ data, and here we concentrate on the multi-band behaviour of this system at higher frequencies. Much less is known for LU Cam apart from its orbital period of 3.6 hours inferred from optical spectroscopy (\citealt{sheets}). We decided to observe LU Cam with ULTRACAM to study in more detail the variability behaviour after one of us (PJG) noted it strong variability on timescales of weeks in the data of the Palomar Transient Factory (\citealt{rau,law}).

In section \ref{sec:data} we will describe our ULTRACAM observations and data reduction procedures to obtain the coherence and phase/time lags. Section \ref{sec:results} will provide the main results and comment on the observed Fourier-dependent features. Our discussions and comparison to similar phenomena observed in XRBs and AGN are presented in section \ref{sec:discussion}, whilst our conclusions are drawn in section \ref{sec:conclusion}.

\section{Observations and data reduction}\label{sec:data}

Both LU Cam and MV Lyr lightcurves presented in this paper were obtained with ULTRACAM (\citealt{ULTRACAM}) at the Cassegrain focus of the 4.2-m William Herschel Telescope (WHT) on La Palma, Spain. ULTRACAM is a CCD camera designed to provide imaging photometry at high temporal resolution in three different filters simultaneously. The instrument provides a 5 arcminute field on its three $1024 \times 1024$ E2V 47-20 CCDs (i.e. 0.3 arcseconds/pixel) on the WHT. Incident light is first collimated and then split into three different beams using a pair of dichroic beamsplitters. For all observations presented here, one beam was dedicated to the SDSS $u'$ ($3543$\AA) filter, another to the SDSS $g'$ ($4770$\AA) filter and the third to the SDSS $r'$ ($6231$\AA) filter. Because ULTRACAM employs frame-transfer chips, the dead-time between exposures is negligible ($\approx 0.02$ seconds). The settings used for each observation can be found in Table \ref{tab:1}, including exposure times and total number of frames. No binning was used and the read-out speed was always set to slow as it provides the lowest detector readout noise ($\approx3$ electrons).
         
\begin{figure*}
\includegraphics[width=0.9\textwidth, height=0.3\textheight]{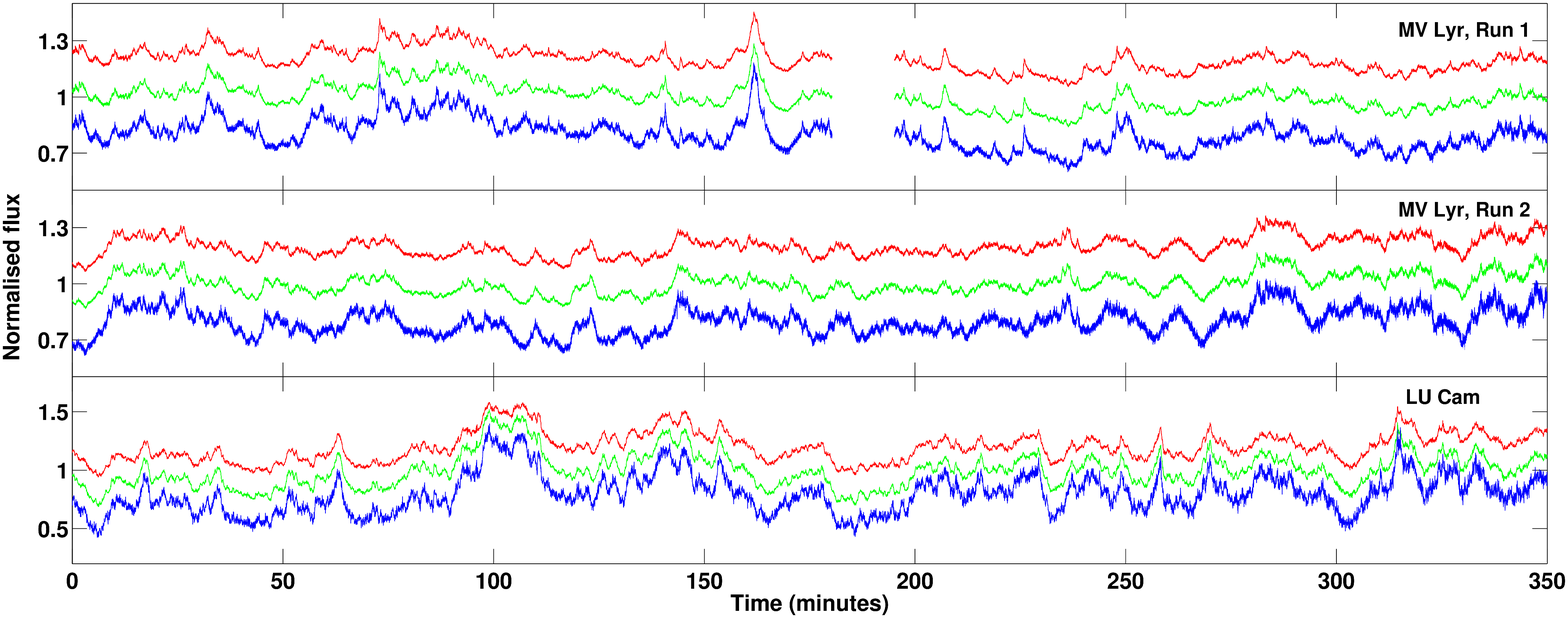}	
\caption{ULTRACAM lightcurves obtained in the $u'$ (bottom-blue), $g'$ (middle-green) and $r'$ (top-red) filters for the CVs MV Lyr (top two panels) and LU Cam (bottom panel). The lightcurves have all been normalised by their respective mean count rate. The $r'$ and $u'$ lightcurves have also been shifted by $0.2$ and $-0.2$ respectively for clarity. Both systems clearly display variability on a wide range of frequencies. Colour figure is available online.}
\label{fig:1}
\end{figure*}

All data were reduced using the ULTRACAM pipeline software. All frames were first debiased and then flat-fielded, the latter using the median of twilight sky frames taken with the telescope spiralling during twilight before observations. For MV Lyr, 2 bright reference stars were used to perform differential photometry, whilst 4 bright comparison stars were used for LU Cam. Furthermore, LU Cam was observed in the $u'$-band in co-add mode. For this mode, one $u'$-band exposure is performed for every two $g'$ and $r'$ exposures to increase signal-to-noise, but nevertheless all three exposures remain synced. To place all filters on the same time stamp, we averaged pairs of the $g'$ and $r'$ exposures. Portions of the extracted lightcurves for MV Lyr and LU Cam in the three filters are shown in Fig.\ref{fig:1}.

\begin{table*}
%\vspace{-19cm}
\centering
\begin{tabular}{l l l l l}
\hline
\hline
Object &  Run  &  Dates    & Exposure/frame (s)   & Frames    \\
\hline
MV Lyr & 1 & 31 August 2012    & 0.844           & 27735 \\
MV Lyr & 2 & 1 September 2012  & 0.526           & 41670 \\
LU Cam & 3 & 10 January 2012   & 2.276           & 18502  \\
\end{tabular}
\caption{ULTRACAM journal of observations during the observing runs.}
\label{tab:1}
\end{table*}

\begin{figure*}
\includegraphics[width=0.9\textwidth]{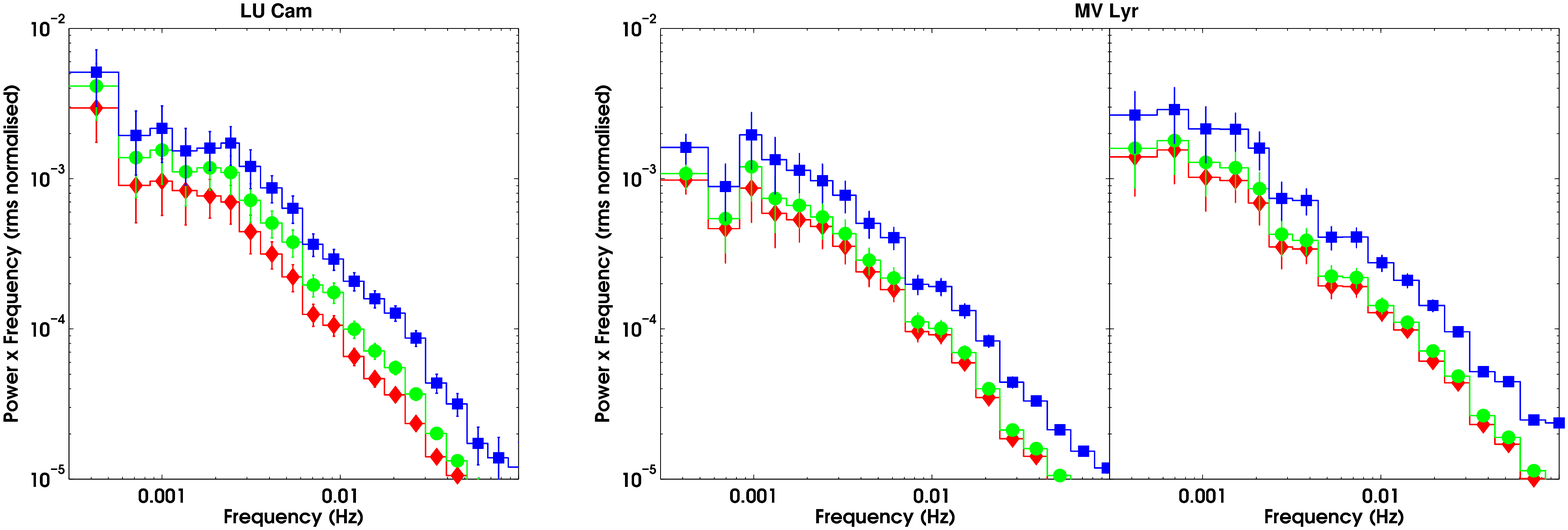}	
\caption{Rms normalised PSDs (power $\times$ frequency with Poisson noise subtracted) for LU Cam (left) and MV Lyr (right, 2 different nights) in the $u'$ (blue squares), $g'$ (green circles) and $r'$ (red diamonds) filters. Each PSD has been obtained by averaging 6 independent PSDs, each obtained from $\approx$1 hour timeseries. Both objects show a flattening of the variability power below $\approx10^{-3}$ Hz. In the case of MV Lyr, this is consistent with the \Kepler\ observations of \protect\cite{scaringi_rms,scaringi_qpo}. Also, both objects display higher levels of variability with increasing photon energy. Colour figure available online.}
\label{fig:2}
\end{figure*}

For each observing run (two for MV Lyr and one for LU Cam) we obtained time averaged power spectral densities (PSDs) by splitting each observing run (each lasting $\approx6$ hours) into 6 non-overlapping segments. For each segment (and for each band) we computed the Fast-Fourier transform (FFT) and applied the rms normalisation of \cite{belloni_hasinger} so that the square root of the integrated PSD power over a specific frequency range yields the root-mean-square (rms) variability. Fig. \ref{fig:2} shows the computed PSDs (in power $\times$ frequency and with Poisson noise subtracted) for all objects and in all bands. All are roughly described with a broken powerlaw up to $\approx10^{-1}$ Hz, after which Poisson noise dominates the PSDs. The PSD break at $\approx10^{-3}$ Hz for MV Lyr is consistent with the results of \cite{scaringi_rms,scaringi_qpo}, and discussed therein.

Additionally, we carried out analysis of cross-spectral Fourier statistics. These are the coherence, phase and time lags, which have been applied to X-ray timing data of XRBs and AGN in the past, and are described in \cite{vaughan} and \cite{nowak}. Because this is the first time such methods are applied to optical data, we here review all the steps involved again, with particular emphasis on optical observations of CVs.   

\subsection{Coherence function}
The coherence function, $\gamma^2$, is a Fourier-frequency-dependent measure of the degree of linear correlation between two simultaneously observed lightcurves in two different filters (or energy bands as used in X-rays). Specifically, it provides the fraction of the mean-squared variability at a specific frequency of one lightcurve which can be directly attributed to the other by a linear transformation (e.g. a simple time shift and flux re-scaling). As a simple intuitive example, two simultaneously observed lighturves in two filters which display a constant time shift relating one to the other will display perfect coherence as well as a constant time lag, at all frequencies.

Let us consider two simultaneously observed lightcurves in two different filters $r(t)$ and $u(t)$. We here use a similar notation to that used in \cite{nowak}, so that $r(t)$ is the longer wavelength band and $u(t)$ the short wavelength band ($s(t)$ and $h(t)$ respectively in \citealt{nowak}). To compute the coherence (and indeed any other higher-order Fourier statistic) we require an ensemble of independent measurements for each band in order to reduce statistical noise. Thus both $r(t)$ and $u(t)$ are split into $M$ independent segments, where $M=6$ in our case for both LU Cam and MV Lyr, so that each segment is about $1$ hour long. For each lightcurve segment, $r_{i=1 \rightarrow M}(t)$ and $u_{i=1 \rightarrow M}(t)$, we compute the Fast Fourier transforms (FFT) $X_{r,i}(f)$ and $X_{u,i}(f)$ respectively, together with the corresponding power spectra $P_{r,i}(f)=|X_{r,i}|^2$ and $P_{u,i}(f)=|X_{u,i}|^2$ and cross-spectra $C_{i}(f)=X_{r,i}^*(f) X_{u,i}(f)$ (where $^*$ denotes the complex-conjugate). 

For each FFT the Poisson (white) noise levels can be calculated as $|N_{r,i}|^2={2 \over \langle{r_{i}(t)\rangle} }$ and $|N_{u,i}|^2={2 \over \langle{u_{i}(t)}\rangle }$ for both bands (using the \cite{belloni_hasinger} normalisation). Because in our case the errors obtained from the differential photometry will not be Poissonian (and could possibly be correlated between the different filter lightcurves due to systematic seeing effects in the comparison stars) we fit each independent PSD with a powerlaw plus a constant to estimate the mean white noise levels $|N_{r,i}|^2$ and $|N_{u,i}|^2$. The intrinsic power for each segment in the $r$ lightcurve is then defined as $|R_i|^2=P_{r,i}-|N_{r,i}|^2$, and similarly for the $u$ lightcurve as $|U_i|^2=P_{u,i}-|N_{u,i}|^2$.

From this, we first define the raw coherence (i.e. the coherence without taking into account white noise) at a specific frequency as
\begin{equation}
\gamma_{raw}^2(f) ={ { |\langle \langle C(f) \rangle \rangle | ^2 \over {\langle \langle |X_{r}(f)|^2 \rangle \rangle} {\langle \langle |X_{u}(f)|^2 \rangle \rangle} } \\ = { |\langle \langle C(f) \rangle \rangle | ^2 \over {\langle \langle P_{r}(f) \rangle \rangle} {\langle \langle P_{u}(f) \rangle \rangle} } }
\label{eq:1}
\end{equation}
where the double angled brackets denote averages over adjacent frequency bins considered ($n_{fs}$ in total) as well as averages between the $M$ independent segments. This notation will be used throughout the paper. The frequency binning in our case is logarithmic, with the constrain that each bin requires at least 5 datapoints, thus $n_{fs}\ge5$ for all bins. The associated error is then defined as 
\begin{equation}
\delta\gamma_{raw}^2(f) = \sqrt{\frac{2}{m}} {\gamma_{raw}^2(f) [1-\gamma_{raw}^2(f)] \over |\gamma_{raw}(f)|},
%\delta\gamma_{raw}^2(f) = \sqrt{\frac{2}{m}} { [1-\gamma_{raw}^2(f)] \over |\gamma_{raw}(f)|},
\label{eq:2}
\end{equation}
where $m=M \times n_{fs}$, the number of measurements used in a specific frequency bin.

The intrinsic coherence, $\gamma^2$, can be estimated by correcting each term in Eq. \ref{eq:1} for counting noise. The term $|\langle \langle C(f) \rangle \rangle |^2$ contains a positive bias (offset) caused by Poisson noise, which can be estimated as
\begin{eqnarray}
n^2 = (\langle \langle |R|^2 \rangle \rangle \langle \langle |N_u|^2 \rangle \rangle + \langle \langle |U|^2 \rangle \rangle \langle \langle |N_r|^2 \rangle \rangle \\ +  \langle \langle |N_r|^2 \rangle \rangle \langle \langle |N_u|^2 \rangle \rangle )/m \nonumber.
\label{eq:3}
\end{eqnarray}
With this, we can define the intrinsic coherence by
\begin{equation}
\gamma^2(f) = { {|\langle \langle C(f) \rangle \rangle | ^2 -n^2}\over {\langle \langle |R(f)|^2 \rangle \rangle} {\langle \langle |U(f)|^2 \rangle \rangle} },
\label{eq:4}
\end{equation}
where the denominator now includes only intrinsic power. The associated error is then defined as 
\begin{eqnarray}
\delta\gamma^2(f) = \gamma^2(f) {1 \over \sqrt{m}} \\ \left[ {2n^4m \over ({|\langle \langle C \rangle \rangle|^2-n^2})^2} +  {\langle \langle |N_r|^4 \rangle \rangle \over \langle \langle |R|^4 \rangle \rangle} + {\langle \langle |N_u|^4 \rangle \rangle \over \langle \langle |U|^4 \rangle \rangle} +{m\delta\gamma_{raw}^4 \over \gamma_{raw}^4} \right]^{1/2} \nonumber .
\label{eq:5}
\end{eqnarray}

\subsection{Fourier phase/time lags}

The Fourier phase/time lags are also computed between two simultaneous lightcurves obtained in different wavelength ranges, and like the coherence function, are related to the cross-correlation (\citealt{bendat,vaughan,nowak}). Similarly to the PSDs in Fig.\ref{fig:2} and the coherence, we split the lightcurves into $M=6$ independent segments. The Fourier phase lag is defined as the phase of the average cross power spectrum, defined as 
\begin{equation}
\Phi(f)=arg[\langle \langle C(f) \rangle \rangle],
\label{eq:6}
\end{equation}
with associated errors defined as
\begin{equation}
\delta\Phi(f)={1 \over \sqrt{m}} \sqrt{{1-\gamma_{raw}^2} \over 2\gamma_{raw}^2}.
\label{eq:7}
\end{equation}

The Fourier time lag is constructed by simply dividing the phase lag by $2\pi f$, so that $\tau(f)={\Phi(f) \over 2\pi f}$. Similarly, the error on the time lag is defined as $\delta\tau(f)={\delta\Phi(f) \over 2\pi f}$.

\section{Results}\label{sec:results}

\begin{figure*}
\includegraphics[width=0.65\textwidth, angle=-90]{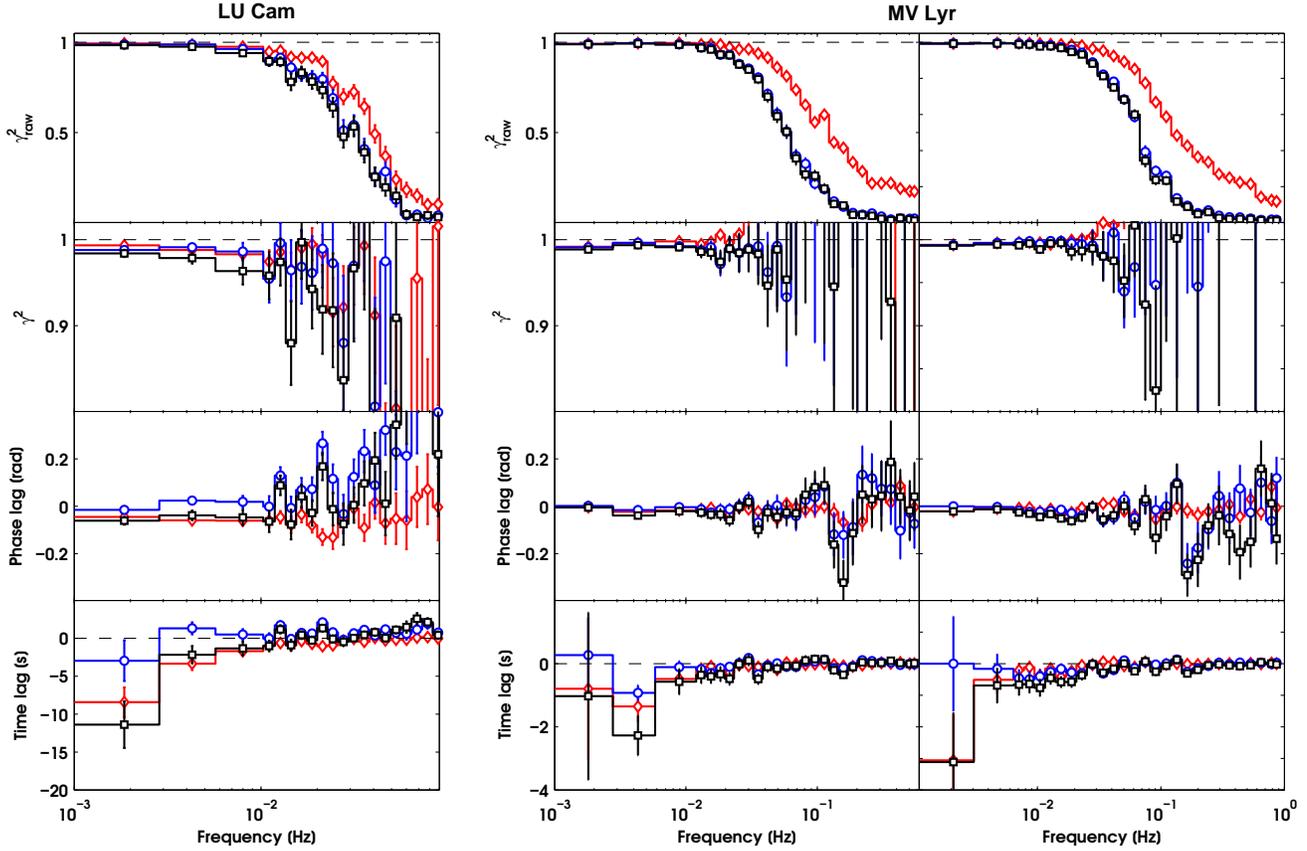}	
\caption{Coherence and phase/time lags observed with ULTRACAM for LU Cam (left) and MV Lyr (right, 2 different nights) in the three combinations $r'-g'$ (red diamonds), $g'-u'$ (blue circles) and $r'-u'$ (black squares). All have been obtained from averaging 6 independent lightcurve segments. Negative time lags indicate that the emission in the redder bands lags the emission in the bluer bands. Colour figure available online.}
\label{fig:3}
\end{figure*}

Fig. \ref{fig:3} shows the coherence functions, phase and time lags for the three observing runs and all three colour combinations $u'-r'$, $u'-g'$ and $g'-r'$. Both LU Cam and MV Lyr show high levels of coherence at the lowest observed frequencies in all three colour combinations. We note that above $\approx10^{-2}$ Hz the noise levels start to affect our observations, resulting in larger error bars in this frequency range. The coherence is a fourth-order statistic and is very sensitive to uncertainties in the noise subtraction (\citealt{bendat,nowak}). 

The noise levels we infer from the PSD fits may also  be slightly over/underestimated for our observations as the errors are not strictly Poissonian. Specifically, we expect errors in adjacent filter lightcurves to be correlated to some degree due systematic effects in the comparison stars. For example, the apparent rise above $1$ in the intrinsic coherence plots of MV Lyr in the $r'-g'$ combination are due to an overestimation of the white noise level. Because of these uncertainties and possible correlations within errors of the different colour combination, the apparently significant features appearing in multiple band selections at high frequencies must be treated with caution. Nevertheless we are confident our coherence and phase/time lag estimates are reliable for frequencies below $\approx10^{-2}$ Hz where uncertainties in the noise levels are negligible compared to the variability power, as also illustrated by the raw coherence plots.

Both LU Cam and MV Lyr also show similar features in the phase/time lags below $\approx10^{-2}$ Hz. Specifically, we detect significant red/negative lags for both systems, where low energy photons are seen arriving after the high energy ones. In the case of LU Cam negative lags are observed with a $\approx$10 second delay below $10^{-3}$ Hz, whilst for MV Lyr this is much smaller at $\approx$3 seconds. Furthermore in both systems the negative lags are most pronounced in the $r'-u'$ combination, which might be expected as this is the filter combination with the largest wavelength difference. At higher frequencies, the time lags in MV Lyr seem to asymptote to $0$, whilst for LU Cam they seem to swap sign reaching a few seconds at the highest observable frequencies. However, because the Poisson noise level in this frequency range affects our observations as described above, the high frequency results must be treated with caution.

\section{Discussion}\label{sec:discussion}

As this is the first study of this kind in CVs, it is non-trivial to predict the Fourier dependent behaviour we have observed. We will thus address our results in the context of different timescales which could potentially explain the observed lags in MV Lyr and LU Cam.

\subsection{Viscous timescale}
If we take the fluctuating accretion disk model, which is frequently invoked to explain the Fourier time lags observed in the X-ray domain for XRBs and AGN (\citealt{uttley11,cassatella1,cassatella2,uttley1,arevalo}), we would expect to observe blue/positive lags in the optical for CVs. This would be a consequence of fluctuations propagating inwards through the disk on the viscous timescale, and emitting redder photons before the blue ones as a consequence of accretion rate fluctuations passing through lower and then higher temperature emitting regions as they move inwards through the disk. What we observe in MV Lyr and LU Cam is the opposite: variations in blue photons are observed preceding the variations in the red photons. Because of this we must rule out this timescale as a possible explanation for the observed lags.

\subsection{Light-travel timescale}
Red/negative lags somewhat similar to those observed here have also been observed in X-rays for XRBs and AGN (referred to as soft/negative lags). For example, \cite{demarco} have shown how soft/negative lags are observed in X-rays for a sample of AGN, up to hundreds of seconds. The main interpretation for these lags is that they represent the light travel time to the disc from the variable continuum source. The continuum source illuminates the disc, which in the case of AGN leads to a soft photoionised reflection spectrum, while in XRBs the absorbed photons are also reprocessed and re-emitted as (soft X-ray) thermal radiation at the local disc blackbody temperature. Thus, in this model, the delay is simply the light-crossing time (plus a small reprocessing time, assuming also that the time-scale for thermal reprocessing is small, see Section \ref{sec:therm}), which for XRBs will be milliseconds, and for AGN will be on the order of tens to hundreds of seconds. This scenario could potentially explain the $\approx1-3$ second lags observed in MV Lyr, placing the reflecting/reprocessing region between 0.4-0.8$R_{\odot}$ (if we take the system parameters from \citealt{linnell,hoard}). This range is both consistent with the outer-edges of the disk and/or the secondary star. 

Although the lag values are consistent with the expected binary separation in MV Lyr, this model would require the $r$-band emission to be dominated by the reprocessed light, otherwise the lags would be diluted by the variable central continuum emission. Furthermore, this interpretation seems to be unlikely for disc reprocessing at these large radii, since the disk is unlikely to reprocess much of the central continuum emission (\citealt{vanpara,dejong}). Additionally, the much larger lags of $\approx10$ seconds observed in LU Cam rule out the possibility of the lag being produced by light-travel times to the outer disc and/or secondary star. LU Cam does however leave the possibility that the reflecting region would sit at large radii, at $\approx6R_{\odot}$, outside of the binary orbit, maybe in the form of circumbinary disk/torus, although we would have to be viewing the system from a very specific angle. If this were the case, we would also expect the large scale varying emission to be weak compared to the direct continuum emission from close to the compact object. This would imply that the observed lags from the reprocessing region at large distances would be diluted from the direct continuum emission, meaning that the intrinsic lags should be even larger than the ones observed in LU Cam.  We note however that for the possibilities presented here further spectral analysis would be required to investigate these possibilities.

\subsection{Thermal timescale} \label{sec:therm}
Another, possibly more plausible, scenario could be that the accretion disk is reprocessing energy originating from close to the compact object, for example in the boundary layer, on the local thermal timescale. In MV Lyr, the boundary layer can reach $\approx 100,000$ K (\citealt{godon}), allowing the colder disk (of $\approx 10,000$ K in the outer edges) to efficiently absorb such photons (as opposed to reflect them through Compton scattering as is the case for XRBs). In this scenario one can imagine variable radiation influencing the accretion disk, where the observed red/negative lags are due to the light-crossing time from the source to the disk, but also due to the thermal reprocessing time of photons within the disk before they are re-emitted and observed. In this case the relevant timescale to consider is the thermal timescale
\begin{equation}
t_{th}={1 \over {\alpha \Omega} },
\label{eq:8}
\end{equation}
where $\Omega$ is the Keplerian frequency at a specific disk radius, and $\alpha$ the disk viscosity parameter, also at a specific disk radius (\citealt{ss_73}). With an accretion disk reaching the WD surface of $\approx 0.01R_{\odot}$, the corresponding thermal timescale at the inner-most edges of the disk is on the order of tens of seconds, depending on $\alpha$. Our LU Cam result could potentially be explained through this process if $\alpha\approx 0.7$, which is considered to be high for accretion disks in CV systems. If on the other hand the reprocessing of photons were to occur at larger radii (say close to the outer edges of the disk at $\approx0.5R_{\odot}$), then the thermal timescale would be much longer. In both scenarios however we would only expect the surface layer of the disk to reprocess photons, greatly reducing the timescale inferred from Eq. \ref{eq:8}, and possibly explaining the observed lags in MV Lyr and LU Cam. We note that the above scenario is a very general example, and we can imagine magnetic fields (instead of photon radiation) affecting the disk, which reprocesses the energy, and emitting photons after some delay, but again comparable to the thermal timescale. The larger red/negative lags observed in AGN are also explained by a similar process, where the inner-disk edge lies at a few AU. The main difference is that for AGN (and XRBs) the X-ray photons originating close to the compact object are re-emitted nearly instantaneously as they photoionise the surface layers of the disk. In CVs on the other hand the optical/UV emission close to the WD cannot photoionise the disk, but instead heats it up, and photons from the disk are then re-emitted on the thermal timescale.

One last possible explanation for the observed lags in LU Cam and MV Lyr is that of reverse (inside-out) shocks within the accretion disk (\citealt{krauland}), possibly originating from the WD boundary layer accretion disk. These waves would then transport energy outwards in the disk. In this scenario the lightcurve variations in the $u$ band will be observed before those in the $r$ band as a consequence of the shock(s) moving outwards in the disk and passing through the hotter inner-edges and later through the cooler outer-edges. The thermal timescale of the disk (\citealt{FKR}) would also be the relevant timescale for the propagation of the shockwave, potentially in line with our observations. This interpretation, as well as all others presented here, require thorough modelling in order to explain the lag phenomena observed in MV Lyr and LU Cam. Such modelling which however beyond the scope of this paper.

\section{Conclusion}\label{sec:conclusion}

We have presented the first analysis of coherence, phase and time lags for CVs at optical wavelengths using ULTRACAM, mounted on the William Herschel Telescope. Our analysis is based on two objects, namely MV Lyr and LU Cam, which both show significant red/negative lags at frequencies below $\approx10^{-2}$, where blue photons are observed before the red ones. For MV Lyr this lag is observed to a maximum of $\approx3$ seconds, whilst for LU Cam the lag is much larger at $\approx10$ seconds. Furthermore we established that both objects show high levels of coherence over a wide range of frequencies up to at least $\approx10^{-2}$ Hz. Mechanisms to explain the observed phenomena have been proposed, involving reprocessing of boundary layer photons (or other sources of energy close to the compact object) from the inner-edge of the accretion disk on the local thermal timescale, or reverse (inside-out) shocks travelling outwards in the disk. A thorough study of the implications of the results presented here, together with detailed modelling of the accretion disk reprocessing is beyond the scope of this Paper. We note however that any models trying to reproduce the observed Fourier-dependent behaviour in XRBs and AGN will, in future, need to address the features presented here as well, in order to provide a self-consistent model explaining the variability behaviour through all wavelengths (including optical) and throughout the different compact accretors (including white dwarfs). Here we have only presented results on two CVs, and noted that both display significant lags at optical wavelengths. We might also expect that most flickering CVs will display similar lags, which will provide further grounds on which to test models seeking to explain the variability properties observed throughout the different types of compact accreting objects. 

\section*{Acknowledgements}
This research has made use of NASA's Astrophysics Data System Bibliographic Services. S.S. acknowledges funding from the FWO Pegasus Marie Curie Fellowship program, as well as funding from NWO project 600.065.140.08N306 to P.J.G. S.S. also wishes to thank Pablo Cassatella for useful and insightful discussions. E.K. acknowledges funding form the NWO Vidi grant 639.042.218.

\bibliographystyle{mn}
\bibliography{lag_paper}

\label{lastpage}

\end{document}